\documentclass[twocolumn,floats,floatfix,superscriptaddress,aps,pra]{revtex4}
\usepackage{amsfonts}
\usepackage{amssymb}
\usepackage{amsmath}
\usepackage{calc}
\usepackage{graphicx}
\usepackage{bm}

\def\be{ \begin{equation} }
\def\ee{ \end{equation} }
\def\bea{ \begin{eqnarray} }
\def\eea{ \end{eqnarray} }
\def\bse{ \begin{subequations} }
\def\ese{ \end{subequations} }

\def\U{\mathbf{U}}

\def\A{\mathcal{A}}

\def\phase{\phi}
\def\th{\vartheta}

\def\i{{\rm{i}}}
\def\f{{\rm{f}}}
\def\e{{\rm{e}}}

\def\to{\rightarrow}

\def\half{\tfrac12}

\def\sec{\section}

\begin{document}

\author{Boyan T. Torosov}
\affiliation{Institute of Solid State Physics, Bulgarian Academy of Sciences, 72 Tsarigradsko chauss\'{e}e, 1784 Sofia, Bulgaria}
\author{Svetoslav S. Ivanov}
\affiliation{Department of Physics, St Kliment Ohridski University of Sofia, 5 James Bourchier blvd, 1164 Sofia, Bulgaria}
\author{Nikolay V. Vitanov}
\affiliation{Department of Physics, St Kliment Ohridski University of Sofia, 5 James Bourchier blvd, 1164 Sofia, Bulgaria}

\title{Narrowband and passband composite pulses for variable rotations}

\date{\today}

\begin{abstract}
We develop a systematic approach to derive narrowband (NB) and passband (PB) composite sequences which can produce any pre-selected transition probability with any desired accuracy.
The NB composite pulses are derived by successive cancellation of derivative terms in the propagator.
The PB pulses are built by combining two half-$\pi$ NB pulses.
Both the NB and PB pulses feature vanishing wings on both sides of the central region around the target transition probability, and the PB pulses offer also a flat (broadband) top around the target probability.
The NB sequences are particularly suitable for sensing and metrology applications, while the PB sequences can be very useful for qubit control in the case of tightly spaced qubits for they ensure both selectivity (suppressed cross talk to neighbors) and robustness (suppressed laser pointing instability).
Other possible applications include selective (NB), or both selective and robust (PB), control of close transitions in frequency space.
\end{abstract}

\maketitle


\sec{Introduction}\label{sec-intro}

Composite pulses have been developed in nuclear magnetic resonance (NMR) over 40 years ago \cite{NMR}. Although the underlying principles and mathematics have been used in polarization optics much earlier \cite{PolarizationOptics}, it is NMR where vast progress on CPs has been achieved.
Due to their advantage of combining the benefits of the accuracy of resonant excitation with a robustness similar to adiabatic techniques, CPs have succeeded to become one of the most popular techniques for control and manipulation of quantum and classical physical systems.
Quite remarkably, CPs allow for a very flexible control of the excitation profile: in addition to broadband (BB), narrowband (NB), and passband (PB) excitation profiles, one can generate virtually any desired excitation shape, which boosts the possibilities for different applications.
During the last years, CPs are undergoing a new wave of intensive development far beyond NMR.
They have been used to improve some well known techniques like rapid adiabatic passage \cite{Torosov2011PRL,Schraft2013}, stimulated Raman adiabatic passage \cite{Torosov2013,Bruns2018}, Ramsey interferometry \cite{Vitanov2015,Zanon-Willette2018}, and dynamical decoupling \cite{Genov2017}.
Applications include qubit control in trapped ions \cite{Gulde2003,Schmidt-Kaler2003,Haffner2008,Timoney2008,Monz2009,Shappert2013,Mount2015}, neutral atoms  \cite{Rakreungdet2009} and doped solids \cite{Schraft2013,Genov2017,Bruns2018}, high-accuracy optical clocks \cite{Zanon-Willette2018}, cold-atoms interferometry \cite{Butts2013,Dunning2014,Berg2015}, optically dense atomic ensembles \cite{Demeter2016}, singlet-triplet quantum-dots qubits \cite{Wang2012,Kestner2013,Wang2014,Zhang2017},
triple quantum dots \cite{Hickman2013,Eng2015}, NV centers in diamond \cite{Rong2015}, magnetometry \cite{Aiello2013},
optomechanics \cite{Ventura2019}, etc.

A composite pulse is a sequence of pulses with different relative phases.
The constituent pulses may have the same or different pulse areas, and may be resonant or possess some detuning and/or chirp.
The relative phases are used as control parameters to shape the excitation in a desired manner.
Usually the profile is shaped by cancellation of derivative terms of the propagator as a function of some chosen parameter, e.g. the Rabi frequency or the detuning. In a recent paper, a different approach to finding the relative phases, based on deep neural network, has been proposed \cite{Yang2018}.

In brief, a single resonant pulse produces a $\sin^2$ excitation profile, which leads to complete population transfer if the pulse area $\A$ is equal to $\pi$, or half excitation if $\A=\pi/2$.
A CP in turn may produce a desired population transfer probability over a whole range of pulse areas if the composite phases are chosen to cancel derivative terms in a Taylor expansion at the point $\A=\pi$ (or $\A=\pi/2$ for half excitation).
In such a way one may cancel errors in one or many experimental parameters.
Recently, composite pulses, termed universal, which cancel systematic errors in any parameter in a two-state system have been also developed \cite{GenovUniversal}.
Finally, CPs that are insensitive to systematic errors in the composite phases have also been derived \cite{TorosovPhaseErrors}.

In a recent work, we developed composite sequences which produce broadband excitation profiles with arbitrary predefined transition probability $P = \sin^2 \theta$ \cite{TorosovBbTheta}, named \emph{broadband theta pulses}.
In the Bloch sphere picture, these CPs move the Bloch vector tip from the north or south pole to any desired parallel of the sphere.
In the CP literature such operations are known as \emph{variable rotations}, as contrasted to \emph{constant rotations}, which move the Bloch vector to any desired point on the Bloch sphere.

In the present work, we extend this idea and show how to produce narrowband and passband variable rotations on the Bloch sphere, which we call \emph{NB and PB $\theta$ pulses}.
NB pulses allow to greatly increase the selectivity of the excitation, which is important in various applications, e.g. in selective spatial addressing of trapped ions or atoms in optical lattices by tightly focused laser beams \cite{SsIvanovLocalAddressing}, or in selective addressing of a particular vibrational sideband frequency mode.
NB $\theta$ pulses can significantly reduce the unwanted cross-talk to other atoms or ions when implementing the Hadamard gate in the quantum register and therefore can increase the fidelity of the quantum circuit.
PB pulses allow to be both selective as NB pulses and robust as BB pulses, at the expense of longer composite sequences.

The paper is organized as follows.
In Sec.~\ref{sec-narrow} we introduce the method for the derivation of the NB composite theta pulses.
The method is based on brute-force numeric cancellation of derivative terms.
Sequences of up to eight pulses have been studied, but longer sequences can be easily found using the same method.
In Sec.~\ref{sec-pass} we show how one can combine two NB half-$\pi$ pulses to produce PB theta pulses.
In Sec.~\ref{sec-compare} we compare our results with relevant CPs in the literature.
Finally, the conclusions and discussion of possible applications are presented in Sec.~\ref{sec-discussion}.

\label{sec-narrow}
\sec{Narrowband pulses}

\begin{table*}
	\begin{tabular}{|c|c|c|c|c|}
		\hline
		$p$ & 2 pulses & 4 pulses  & 6 pulses & 8 pulses \\
		{} & $A_{0} A_{\phi_2}$ &
		$A_{0} B_{\phi_2} B_{\phi_3} A_{\phi_4}$ &
		$A_{0} B_{\phi_2} B_{\phi_3} B_{\phi_4} B_{\phi_5} A_{\phi_6}$ &
		$A_{0} B_{\phi_2} B_{\phi_3} B_{\phi_4} B_{\phi_5}  B_{\phi_6} B_{\phi_7}A_{\phi_8}$\\
		\hline
		{} & $\phi_2$  & $\phi_2,\phi_3,\phi_4$ &  $\phi_2,\phi_3,\phi_4,\phi_5,\phi_6$  &  $\phi_2,\phi_3,\phi_4,\phi_5,\phi_6,\phi_7,\phi_8$  \\
		\hline
		$0.1$ & $0.7952 $ & $0.0769, 1.0257, 1.1026$  & $1.4150, 0.5716, 0.8499, 0.0064, 1.4214 $ & $1.2681, 0.5191, 0.4643, 1.5937, 1.5389, 0.7899, 0.0580$  \\
		$0.2$ & $0.7048 $ & $0.1108, 1.0373, 1.1481$  & $1.4316, 0.6075, 0.8012, 1.9772, 1.4087 $ & $1.2813, 0.5427, 0.4539, 1.6112, 1.5223, 0.7838, 0.0651$  \\
		$0.25$& $0.6667 $ & $0.1252, 1.0422, 1.1674$  & $1.4355, 0.6191, 0.7820, 1.9656, 1.4011 $ & $1.2851, 0.5505, 0.4482, 1.6161, 1.5138, 0.7792, 0.0643$  \\
		$0.3$ & $0.6310 $ & $0.1386, 1.0469, 1.1855$  & $1.4379, 0.6284, 0.7646, 1.9551, 1.3930 $ & $1.2879, 0.5569, 0.4423, 1.6198, 1.5052, 0.7742, 0.0621$  \\
		$0.4$ & $0.5641 $ & $0.1639, 1.0557, 1.2196$  & $1.4400, 0.6430, 0.7330, 1.9360, 1.3760 $ & $1.2917, 0.5672, 0.4302, 1.6248, 1.4879, 0.7633, 0.0551$ \\
		$0.5$ & $0.5 $    & $0.1881, 1.0644, 1.2525$  & $1.4396, 0.6541, 0.7038, 1.9182, 1.3579 $ & $1.2939, 0.5752, 0.4177, 1.6277, 1.4702, 0.7515, 0.0454$ \\
		$0.6$ & $0.4359 $ & $0.2124, 1.0732, 1.2857$  & $1.4374, 0.6629, 0.6752, 1.9008, 1.3382 $ & $1.2948, 0.5818, 0.4043, 1.6291, 1.4516, 0.7386, 0.0334$ \\
		$0.7$ & $0.3690 $ & $0.2379, 1.0827, 1.3207$  & $1.4334, 0.6702, 0.6460, 1.8828, 1.3162 $ & $1.2947, 0.5874, 0.3896, 1.6291, 1.4314, 0.7241, 0.0187$ \\
		$0.75$& $0.3333 $ & $0.2515, 1.0879, 1.3395$  & $1.4307, 0.6734, 0.6306, 1.8732, 1.3039 $ & $1.2942, 0.5899, 0.3815, 1.6286, 1.4202, 0.7159, 0.0101$  \\
		$0.8$ & $0.2952 $ & $0.2661, 1.0936, 1.3597$  & $1.4274, 0.6763, 0.6142, 1.8630, 1.2904 $ & $1.2934, 0.5922, 0.3727, 1.6277, 1.4081, 0.7069, 0.0003$ \\
		$0.9$ & $0.2048 $ & $0.3009, 1.1075, 1.4083$  & $1.4183, 0.6813, 0.5755, 1.8385, 1.2568 $ & $1.2906, 0.5965, 0.3508, 1.6240, 1.3784, 0.6843, 1.9749$ \\
		$1.0$ & $0$       & $0.3807, 1.1422, 1.5229$  & $1.3915, 0.6844, 0.4873, 1.7802, 1.1717$  & $1.2794, 0.6001, 0.2971, 1.6086, 1.3056, 0.6262, 1.9057$ \\
		\hline
	\end{tabular}
	\caption{
		Phases of composite pulse sequences which produce NB profiles with different transition probability.
		All phases are given in units $\pi$.
	}
	\label{Table:primes}
\end{table*}

We derive the NB $\theta$ pulses in a way similar to our previous works \cite{Torosov2011PRA, Torosov2011PRL, TorosovTwin, TorosovBbTheta}. We provide a brief description below.
The propagator of a coherently driven two-state system can be written as
\be\label{SU(2)}
\U_0 = \left[ \begin{array}{cc} a & b \\ -b^{\ast} & a^{\ast} \end{array}\right],
\ee
where $a$ and $b$ are  the Cayley-Klein parameters and $|a|^2+|b|^2=1$.
For exact resonance ($\Delta=0$), which we assume in this work, $a=\cos(\A/2) $, $b=-\i\sin(\A/2)$, where $\A$ is the temporal pulse area $\A=\int_{t_\i}^{t_\f}\Omega(t)\mathrm{d} t$. A phase shift in the driving field $\Omega\to\Omega\e^{\i\phi}$ is mapped into the propagator as
\be\label{U phase}
\U_0\to \U_\phase = \left[ \begin{array}{cc} a & b \e^{\i\phase} \\ -b^{\ast}\e^{-\i\phase} & a^{\ast} \end{array}\right].
\ee
A train of $N$ pulses, each with area $\A_k$ and phase $\phase_k$,
\be\label{CPsequence}
(\A_1)_{\phase_1} (\A_2)_{\phase_2} (\A_3)_{\phase_3} \cdots (\A_N)_{\phase_{N}},
\ee
produces the propagator
\be\label{U^N}
\U^{(N)} = \U_{\phase_{N}}(\A_N) \cdots \U_{\phase_{3}}(\A_3) \U_{\phase_{2}}(\A_2) \U_{\phase_{1}}(\A_1).
\ee

\begin{figure}[tb]
	\includegraphics[width=6.5cm]{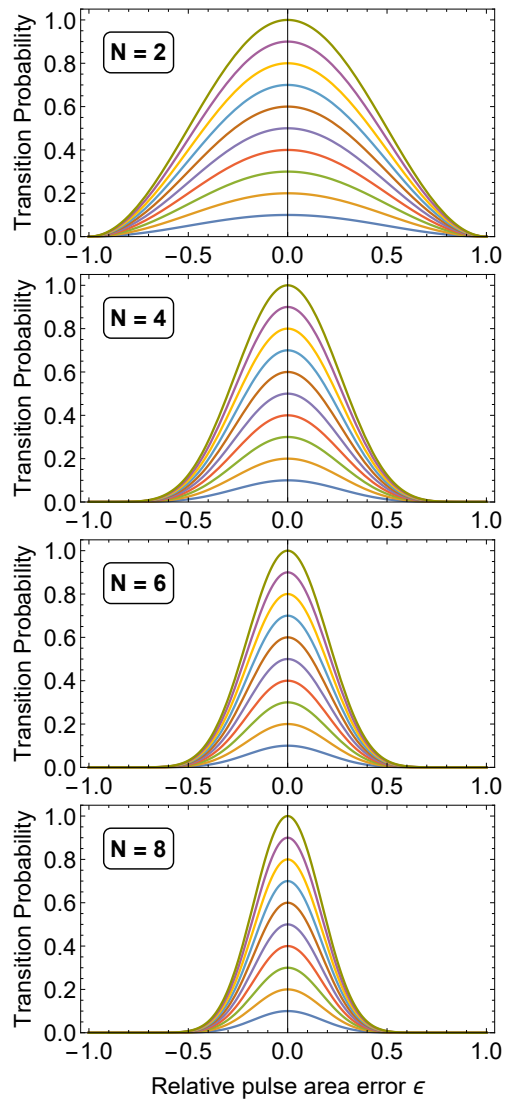}
	\caption{Transition probabilities induced by composite narrowband theta pulses of different length $N$, with phases given in Table~\ref{Table:primes}. The curves in each frame show the transition probabilities locked at the levels $0.1, 0.2,\ldots, 1.0$.}
	\label{NB2468}
\end{figure}

In this paper, we consider composite sequences of the type
\be\label{NbSequence}
A_{\phase_1} B_{\phase_2}B_{\phase_3}\cdots B_{\phase_{N-1}}A_{\phase_N} ,
\ee
where $A=\pi(1+\epsilon)/2$ is a nominal (for zero error, $\epsilon=0$) $\pi/2$ pulse and $B=2A =\pi(1+\epsilon)$ is a nominal $\pi$ pulse. Furthermore, due to the irrelevance of the global phase, we set $\phi_1=0$. If we consider the total propagator as a function of $\epsilon$, $\U^{(N)}=\U^{(N)}(\epsilon)$, our goal is to choose the phases $\phase_k$ such that we cancel as many derivative terms of the type
\be\label{derivatives}
s_k = \left.\frac{\partial^k}{\partial\epsilon^k}U_{12}^{(N)}\right|_{\epsilon=1}
\ee
as possible, while keeping the transition probability at $\epsilon=0$ to a predefined value $p$,
\be\label{zeroOrder}
\left|U_{12}^{(N)}(0)\right|^2 = p.
\ee
It can be shown that due to the symmetry in Eq.~\eqref{NbSequence}, the even-order derivative terms in Eq.~\eqref{derivatives} cancel out, $s_{2j}=0$.
Due to this fact, we will stick to an even total number of pulses $N$ in our sequences, as any chosen accuracy, provided by an odd-$N$ sequence, can be achieved by an even $(N-1)$ sequence.
For even $N$, condition \eqref{zeroOrder} imposes the relation
\be\label{relation}
\cos^2 \left(\sum_{k=2}^{N-1} (-1)^k \phi_k + \frac{\phi_N}{2}\right) = p.
\ee

For $N=2$, the transition probability reads
\be\label{P2}
P = \cos ^2\left(\frac{\pi  \epsilon }{2}\right) \cos ^2\left(\frac{\phi _2}{2} \right)
\ee
The phase $\phi_2$ is obtained immediately from here or from Eq.~\eqref{relation},
\be
\phi_2 = 2 \arccos\left(\sqrt{p}\right).
\ee
The values of $\phi_2$ for a set of values of the transition probability are given in Table \ref{Table:primes}.

For $N=4$, we arrive at the equations
\bse
\begin{align}
\cos^2(\phi_2-\phi_3+\phi_4/2) &= p,\\
1 + 2e^{i\phi_2}+ 2e^{i\phi_3}+ e^{i\phi_4} &= 0.
\end{align}
\ese
One of the solutions reads
\bse
\begin{align}
\phi_2 &= 2 \arccos\left(\sqrt{p^\prime}\right), \\
\phi_3 &= \phi_4 -\phi_2, \\
\phi_4 &= \pi + 2\arg(1+2e^{i\phi_2}),
\end{align}
\ese
where
\be
\begin{split}
p^\prime = \tfrac{1}{4} &\left[\sqrt{\left(1-p^{\frac13}\right) \left(2 \sqrt{p^{\frac23}+p^{\frac13}+1}+p^{\frac13}+2\right)}\right. \\
&\left. +\sqrt{p^{\frac23}+p^{\frac13}+1}+1\right].
\end{split}
\ee
The values of these phases for a set of values of the transition probability are given in Table \ref{Table:primes}.

For $N \geq 6$, $\phase_k$ are obtained numerically and their values are given in Table~\ref{Table:primes}.

The excitation profile, produced by these composite sequences, reads
\be\label{P-NB}
P = p\cos^{2(N-1)}\left(\frac{\pi\epsilon}{2}\right) .
\ee
Several important conclusions follow from this simple formula.
First, the excitation profile is symmetric, with a maximum value of $p$ at $\epsilon=0$, and vanishing wings toward $\epsilon=\pm 1$.
Second, formula \eqref{P-NB} can be used to explicitly evaluate the width (half-width at half-maximum) of the NB excitation profile: $\epsilon_{\frac12} = \arccos(2^{\frac{N-2}{N-1}}-1)/\pi$.
For NB composite sequences of $N=2,4,6,8$ pulses this gives $\epsilon_{\frac12} = 0.5$, 0.3, 0.234, 0.199.
Third,  formula \eqref{P-NB} shows the NB suppression order at $\epsilon = \pm 1$: $O((\epsilon \mp 1)^{2(N-1)})$ for a NB sequence of $2N$ pulses.
All features described above indicate that the more pulses we have in our sequence, the narrower the excitation profile is.
This can be seen from Fig.~\ref{NB2468}, where we plot the transition probability of the NB theta pulses for different values of the central probability $p$.

It is very important to note that in addition to the excitation suppression in the wings of the excitation profile, the NB composite sequences feature a smooth maximum of value $p$ at $\epsilon=0$.
Indeed, as evident from Eq.~\eqref{P-NB}, for small $\epsilon$ we have $P = p [1 - (N-1) (\pi\epsilon/2)^2 + O(\epsilon^4)]$.
This means that at the target probability $p$ the NB pulses are robust up to order $O(\epsilon^2)$ to variations of $\epsilon$. [A single resonant pulse is robust only to the first order, $O(\epsilon)$, except for $\A=\pi$.]
The robustness to variations in $\epsilon$ can be firther boosted by PB pulses, which are described below.

\label{sec-pass}
\sec{Passband pulses}

\begin{figure}[tb]
	\includegraphics[width=6.5cm]{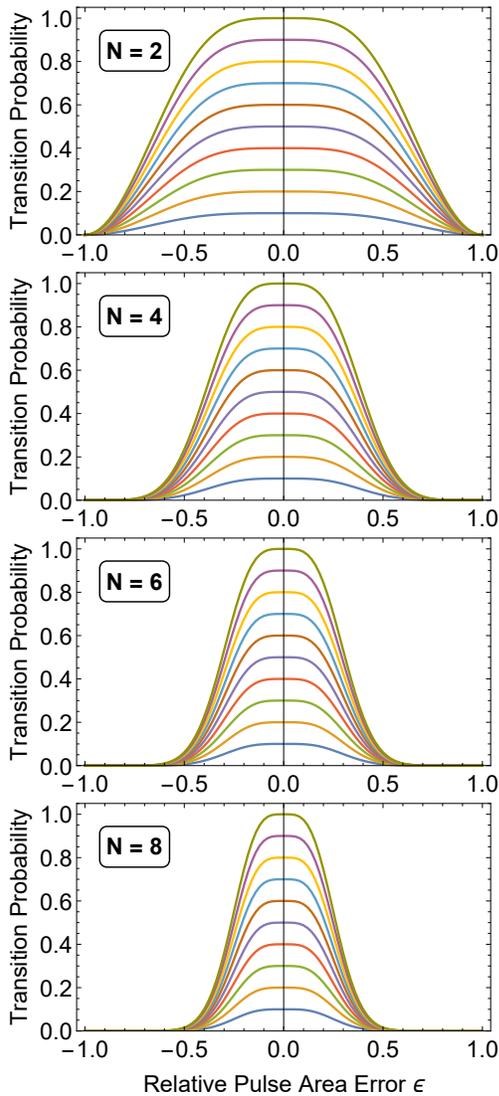}
	\caption{Transition probabilities induced by composite PB theta pulses of different length $2N$. The curves in each frame show the transition probabilities locked at the levels $0.1, 0.2,\ldots, 1.0$.}
	\label{PB2468}
\end{figure}

We can use the idea from Ref.~\cite{TorosovBbTheta} to produce PB theta pulses by ``twinning'' two NB $\pi/2$ pulses.
We consider the pulse sequence \eqref{NbSequence}, followed by the reversed pulse sequence, shifted with some phase $\th$,
\be\label{half-pi-sym-reversed-theta}
A_{\phi_N+\th} B_{\phi_{N-1}+\th} \cdots B_{\phi_3+\th} B_{\phi_2+\th} A_{\phi_1+\th}.
\ee
As shown in \cite{TorosovBbTheta}, the excitation profile for the total pulse sequence is given by
\be
P = 4p_s(1-p_s)\cos^2(\half\th),
\ee
where $p_s$ is the single (non-twinned) transition probability of the pulse sequence.
Therefore, when $p_s=\half$, the total transition probability is determined by the phase $\th$.
If we set this phase to the value $\th=2\arccos(\sqrt{p})$, we obtain the desired transition probability $P=p$.
If the composite $\pi/2$ sequences are accurate to order $O(\epsilon^N)$, i.e. $p_s=\half + c\epsilon^N$, then the composite $\theta$ pulse will be accurate to order $O(\epsilon^{2N})$, since $4p_s(1-p_s)=1-4c^2\epsilon^{2N}$.
Therefore the total sequence, built by two $\pi/2$ NB pulses, will have a broadband profile in its centre.
In the wings, where $p_s=c\epsilon^N$, we have $4p_s(1-p_s)=4c\epsilon^{N}-4c^2\epsilon^{2N}$, which means that the NB property is conserved up to the same order.
Therefore, by twinning two NB  half-$\pi$ pulses, we can obtain a PB CP with any desired central transition probability $p$.
We show the profiles of these PB CPs in Fig.~\ref{PB2468}, for different total number $2N$ of constituent pulses.
The comparison of Fig.~\ref{PB2468} to Fig.~\ref{NB2468} demonstrates that the PB profiles maintain the NB feature of the NB profiles, however, with the added benefit of a flat top.
Therefore, the PB composite sequences feature both selectivity and robustness, at the expense of being a factor of 2 longer than the NB sequences.

\sec{Comparison with other composite pulses}\label{sec-compare}

\begin{figure}[tb]
	\includegraphics[width=6.5cm]{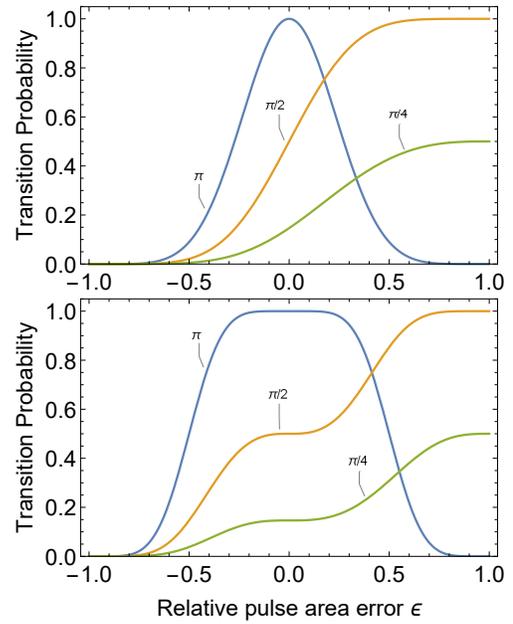}
	\caption{Transition probabilities for the NB (top) and PB (bottom) pulses of Ref.~\cite{Wimperis1994}. The curves in each frame show the transition probabilities for $\theta=\pi,\pi/2,\pi/4$, which correspond to probability levels 0.25, 0.5, and 1.}
	\label{WimperisFig}
\end{figure}

We compare our sequences with the NB and PB sequences of Wimperis \cite{Wimperis1994, Wimperis1989}, which are the most prominent NB and PB pulses with arbitrary transition probability (flip angle).
The NB sequences in \cite{Wimperis1994} are given by
\be
B_{\phi}C_{-\phi}B_{\phi}\theta_0 ,
\ee
where $C=2B = 2\pi(1+\epsilon)$ is a nominal $2\pi$ pulse, $\theta=2\arcsin(\sqrt{p})$, and $\phi=\arccos(-\theta/4\pi)$. The PB sequences are
\be
C_{\chi}C_{-\chi}C_{-\chi}C_{\chi}\theta_0 ,
\ee
where $\chi=\arccos(-\theta/8\pi)$.
In Fig.~\ref{WimperisFig} we show the excitation profiles of the NB and PB pulses, derived in Ref.~\cite{Wimperis1994}.
As one can see from the figure, the profiles of these pulses are asymmetric (except for $p=1$).
This is a drawback if one strives for an excitation limited only to a certain region and no excitation on either sides.
In Ref.~\cite{Wimperis1989} a similar PB half-$\pi$ CP is derived, which yields the same profile as in \cite{Wimperis1994} with smaller pulse area. However, it also exhibits an asymmetric excitation profile.
Finally, NB and PB pulses have been derived in the context of high-fidelity individual addressing of closely spaced particles \cite{SsIvanovLocalAddressing}.
Again, these CPs yield asymmetric profiles since they were specifically designed to robustly manipulate a particle while leaving its neighbors unaffected.
Such a setup requires one to suppress only transitions driven by relatively small pulse areas ``seen'' by the neighboring particles.

\sec{Discussion and Conclusions}\label{sec-discussion}

In this work we presented a class of composite pulses which produce NB and PB excitation profiles at any desired pre-selected transition probability.
Contrary to most NB and PB CPs for arbitrary rotations published in the literature the present CPs feature vanishing wings on both sides of the central region around the target transition probability.
In addition to the NB profiles, this NB feature is present in the PB profiles too where it is accompanied by a BB top around the target probability.
This makes the NB sequences particularly appropriate for sensing applications, while the PB sequences can be very useful for qubit control in the case of closely spaced qubits, e.g. in 1D or 2D structures of trapped ions, or ultracold atoms in optical lattices.
In such situations both selectivity (suppressed unwanted cross talk to neighbors) and robustness (suppressed laser pointing instability) are highly desirable.

Other applications include selective (NB), or both selective and robust (PB), control of close transitions in frequency space.
For example, this necessity emerges when driving vibrational sidebands in trapped ions, as the couplings (and hence the Rabi frequency and the pulse area) depend on the vibrational quantum number: hence NB and PB CPs provide the possibility for very selective addressing.
Another example is molecular chiral resolution.
Recently, it has been shown that a robust and high-fidelity enantio-sensitive population transfer can be achieved by using appropriate BB composite sequences \cite{TorosovDrewsenVitanov}.
By replacing the BB pulses with PB or NB analogues, depending on the studied molecule, one could improve the method even further by their greater selectivity.
Finally, the presented CPs may find useful applications related to sensing, high-precision spectroscopy, optical elements, etc.

\acknowledgments
This work is supported by the European Commission's Horizon-2020 Flagship on Quantum Technologies project 820314 (MicroQC).


\end{document}